\documentclass[twocolumn,citeautoscript,showpacs,preprintnumbers,amsmath,amssymb,prb,floatfix,footinbib]{revtex4}
\usepackage{graphicx}% Include figure files
\usepackage{dcolumn}% Align table columns on decimal point
\usepackage{bm}% bold math
\usepackage{times}% font that prx use
\begin{document}
\title{Character of the structural and magnetic phase transitions in the parent and electron doped BaFe$_{2}$As$_{2}$ compounds}
\author{M. G. Kim\textsuperscript{1}, R. M. Fernandes\textsuperscript{1}, A. Kreyssig\textsuperscript{1}, J. W. Kim\textsuperscript{2}, A. Thaler\textsuperscript{1}, S. L. Bud'ko\textsuperscript{1}, P. C. Canfield\textsuperscript{1}, R. J. McQueeney\textsuperscript{1}, J. Schmalian\textsuperscript{1} and A. I. Goldman\textsuperscript{1}}
\affiliation{\\\textsuperscript{1}Ames Laboratory, U.S. DOE and Department of Physics and Astronomy\\
Iowa State University, Ames, IA 50011, USA}
\affiliation{\\\textsuperscript{2}Advanced Photon Source, Argonne
National Laboratory, Argonne, Illinois 60439, USA}

\date{\today}

\begin{abstract}
We present a combined high-resolution x-ray diffraction and x-ray
resonant magnetic scattering (XRMS) study of as-grown
BaFe$_2$As$_2$. The structural/magnetic transitions must be
described as a two-step process. At $T_S$ = 134.5 K we observe the
onset of a second-order structural transition from the
high-temperature paramagnetic tetragonal structure to a paramagnetic
orthorhombic phase, followed by a discontinuous step in the
structural order parameter that is coincident with a first-order
antiferromagnetic (AFM) transition at $T_N$ = 133.75 K.  These data,
together with detailed high-resolution x-ray studies of the
structural transition in lightly doped
Ba(Fe$_{1-x}$Co$_{x}$)$_2$As$_2$ and
Ba(Fe$_{1-x}$Rh$_{x}$)$_2$As$_2$ compounds, show that the structural
and AFM transitions do, in fact, occur at slightly different
temperatures in the parent BaFe$_2$As$_2$ compound, and evolve
towards split second-order transitions as the doping concentration
is increased. We estimate the composition for the tricritical point
for Co-doping and employ a mean-field approach to show that our
measurements can be explained by the inclusion of an anharmonic term
in the elastic free energy and magneto-elastic coupling in the form
of an emergent Ising-nematic degree of freedom.
\end{abstract}


\maketitle
\section{Introduction}
Although most of the excitement surrounding the
discovery\cite{kamihara_iron-based_2008,rotter_superconductivity_2008}
of superconductivity (SC) in the iron pnictide compounds has focused
on the underlying mechanism, these compounds also provide us with a
fascinating opportunity to study subtle, and not so subtle, aspects
of the interactions between SC, structure and magnetism.  A great
deal of this research has focused on the $\emph{R}$FeAsO ($\emph{R}$
= rare earth), or \emph{1111} compounds, and the
$\emph{AE}$Fe$_2$As$_2$ ($\emph{AE}$ = alkaline earth), or
\emph{122} family. The parent \emph{1111} and \emph{122} systems
undergo transitions from a high-temperature paramagnetic (PM)
tetragonal phase to a low-temperature antiferromagnetic (AFM)
orthorhombic
structure\cite{lumsden_magnetism_2010,rotter_spin-density-wave_2008,goldman_lattice_2008,yan_structural_2008,zhao_spin_2008,jesche_strong_2008,huang_neutron-diffraction_2008}.
In the \emph{1111} parent compounds, the structural transition
occurs at somewhat higher temperature than the AFM
ordering.\cite{lumsden_magnetism_2010} In the \emph{122} parent
compounds, these transitions appear coincident in temperature and
coupled.\cite{goldman_lattice_2008,zhao_spin_2008,jesche_strong_2008,huang_neutron-diffraction_2008}

The consequences of electron/hole doping for the BaFe$_2$As$_2$
compound have been studied widely. For example, upon electron doping
through substitution on the Fe site by Co, Ni, Rh, Pt or Pd, the
structural and AFM transitions are suppressed and split, with the
structural transition found at somewhat higher
temperature.\cite{canfield_feas_2010,ni_effects_2008,Chu_2009,li_superconductivity_2009,Lester_2009,pratt_coexistence_2009,Christianson_2009,kreyssig_suppression_2010,wang_electron-doping_2010}
Superconductivity emerges over a small but finite range in dopant
concentration. In contrast, the structural and AFM transitions
remain coincident in superconducting samples produced by K doping on
the Ba site, P doping on the As site, and Ru doping on the Fe
site.\cite{rotter_superconductivity_2008,chen_coexistence_2009,jiang_superconductivity_2009,thaler_physical_2010,kim_neutron_2010}
In Co-, Rh-, Ni-, and Ru-doped BaFe$_2$As$_2$, neutron and x-ray
measurements revealed a suppression in both the magnetic order
parameter and the orthorhombic distortion below the superconducting
transition temperature, which indicates the coexistence and
competition between magnetism and superconductivity in these
systems.\cite{pratt_coexistence_2009,Christianson_2009,kreyssig_suppression_2010,wang_electron-doping_2010,nandi_anomalous_2010,kim_neutron_2010,fernandes_unconventional_2010}

The nature of the structural and AFM transitions has, itself, been
the subject of intense scrutiny.  For CaFe$_2$As$_2$ and
SrFe$_2$As$_2$, several neutron, x-ray, $\mu$SR and NMR measurements
have reported that the structural/magnetic transitions are
discontinuous (first-order) and
hysteretic.\cite{zhao_spin_2008,goldman_lattice_2008,yan_structural_2008,jesche_strong_2008,kitagawa_antiferromagnetism_2009}
However, in BaFe$_2$As$_2$, there has been significant debate
concerning the character of the structural and magnetic order
parameters. Early neutron measurements on polycrystalline samples
found a second-order magnetic
transition\cite{huang_neutron-diffraction_2008}. Neutron diffraction
measurements by some groups, however, described a first-order
magnetic transition but failed to observe any
hysteresis,\cite{matan_anisotropic_2009} while neutron and NMR
measurements by others found a first-order magnetic transition and
observed a large hysteresis upon cooling and
warming\cite{kofu_title_2009,kitagawa_commensurate_2008}. Wilson
\emph{et al.} initially reported that the magnetic and structural
transitions in BaFe$_2$As$_2$ were continuous in nature, and could
be described by a simple power-law dependence with an critical
exponent consistent with a two-dimensional Ising
model.\cite{wilson_neutron_2009,wilson_2010_1} Later heat capacity
and x-ray work by this group on annealed samples of BaFe$_2$As$_2$
found that the orthorhombic distortion appeared first as a
second-order transition interrupted, at slightly lower temperature,
by a first-order transition to the low temperature orthorhombic
phase.\cite{Rotundu_2010}  As noted by these authors, this complex
structural transition and its relationship to the concomitant AFM
ordering calls for further investigations.

To shed light on the nature of structural and magnetic transitions,
we present a combined high-resolution x-ray diffraction and x-ray
resonant magnetic scattering (XRMS) study of as-grown
BaFe$_2$As$_2$.  In Section II we provide details of the scattering
experiments on the parent BaFe$_2$As$_2$ compound and the Co- and
Rh-doped compounds. In Sections III, we report the results of
high-resolution x-ray diffraction and XRMS measurements on the
parent BaFe$_2$As$_2$ compound and show that there is a small, but
distinct, difference in the temperatures of the structural and AFM
transitions.  We also studied the doped compounds to substantiate
our conclusion that the structural transition is continuous in
nature, whereas the AFM transition changes from a first-order
transition at low doping to a second-order transition for higher
doping levels as the system passes through a tricritical point.  In
Section IV, we estimate the position of this tricritical point for
the Ba(Fe$_{1-x}$Co$_{x}$)$_2$As$_2$ phase diagram and employ a
mean-field approach to show that our measurements can be explained
by the inclusion of an anharmonic term in the elastic free energy
and magneto-elastic coupling in the form of an emergent
Ising-nematic degree of freedom.  In Section V, we summarize our
results.

\section{Experimental Details}
Single crystals of BaFe$_2$As$_2$, Ba(Fe$_{1-x}$Co$_{x}$)$_2$As$_2$
and Ba(Fe$_{1-x}$Rh$_{x}$)$_2$As$_2$ compounds were produced using
the self-flux solution growth method described
elsewhere.\cite{ni_effects_2008} Energy dispersive spectroscopy
(EDS) was performed to confirm the absence of foreign elements, and
wavelength dispersive spectroscopy (WDS) was employed to determine
the compositions of the Co- and Rh-doped compounds at several points
on each sample, providing a combined statistical and systematic
error of less than 5$\%$ of the relative elemental concentration
[e.g. 0.018$\pm$0.001 for the Ba(Fe$_{0.982}$Co$_{0.018}$)$_2$As$_2$
sample]. Temperature-dependent ac electrical resistance data
(\emph{f} = 16 Hz, \emph{I} = 3 mA) were collected using a Quantum
Design Magnetic Properties Measurement System (MPMS) with a Linear
Research LR700 resistance bridge. Electrical contact was made to the
sample using Epotek H20E silver epoxy to attach Pt wires in a
four-probe configuration.

Temperature-dependent, high-resolution, single-crystal x-ray
diffraction measurements were performed on a four-circle
diffractometer using Cu \emph{K}$_{\alpha1}$ radiation from a
rotating anode x-ray source, selected by a germanium (1~1~1)
monochromator. For these measurements, the plate-like single
crystals with typical dimensions of
3\,$\times$\,3\,$\times$\,0.5\,mm$^3$ were attached to a flat copper
sample holder on the cold finger of a closed-cycle displex
refrigerator. The mosaicities of the BaFe$_2$As$_2$,
Ba(Fe$_{1-x}$Co$_x$)$_2$As$_2$ and Ba(Fe$_{1-x}$Rh$_x$)$_2$As$_2$
single crystals were all less than 0.02$^\circ$
full-width-at-half-maximum (FWHM) as measured by the rocking curves
of the (1~1~10) reflection at room temperature. The diffraction data
were obtained as a function of temperature between room temperature
and 8~K, the base temperature of the refrigerator.

To correlate the evolution of the structure with the occurrence of
magnetic order, both conventional x-ray diffraction and XRMS
measurements were performed on the 6ID-B beamline at the Advanced
Photon Source (APS) using the same as-grown BaFe$_2$As$_2$ single
crystal studied with the laboratory source. The single crystal was
attached to a flat copper sample holder on the cold finger of a
closed-cycle displex refrigerator with the tetragonal ($HHL$) plane
coincident with the vertical scattering plane.  The temperature,
measured at a sensor mounted to the copper block holding the sample,
was stable within $\pm$0.002 K. Care was taken to ensure that
heating effects associated with the incident x-ray beam were
minimized by measuring charge and magnetic reflections in close
proximity and using the appropriate incident beam attenuation.
Measurements of both the charge scattering and the XRMS were done at
the Fe $K$-edge ($E$ = 7.112 keV).\cite{kim_resonant_2010} The
incident radiation was linearly polarized perpendicular to the
vertical scattering plane ($\sigma$-polarized) with a spatial cross
section of 1.0 mm (horizontal) $\times$ 0.2 mm (vertical). In this
configuration, dipole resonant magnetic scattering rotates the plane
of linear polarization into the scattering plane
($\pi$-polarization), while the charge scattering leaves the
polarization unchanged. Cu (2 2 0) was used as a polarization
analyzer to suppress the charge and fluorescence background relative
to the magnetic scattering signal by approximately a factor of 200.

\section{Results}
\subsection{High-resolution x-ray diffraction and XRMS measurements of BaFe$_2$As$_2$}
\begin{figure}
\begin{center}
\includegraphics[clip, width=.42\textwidth]{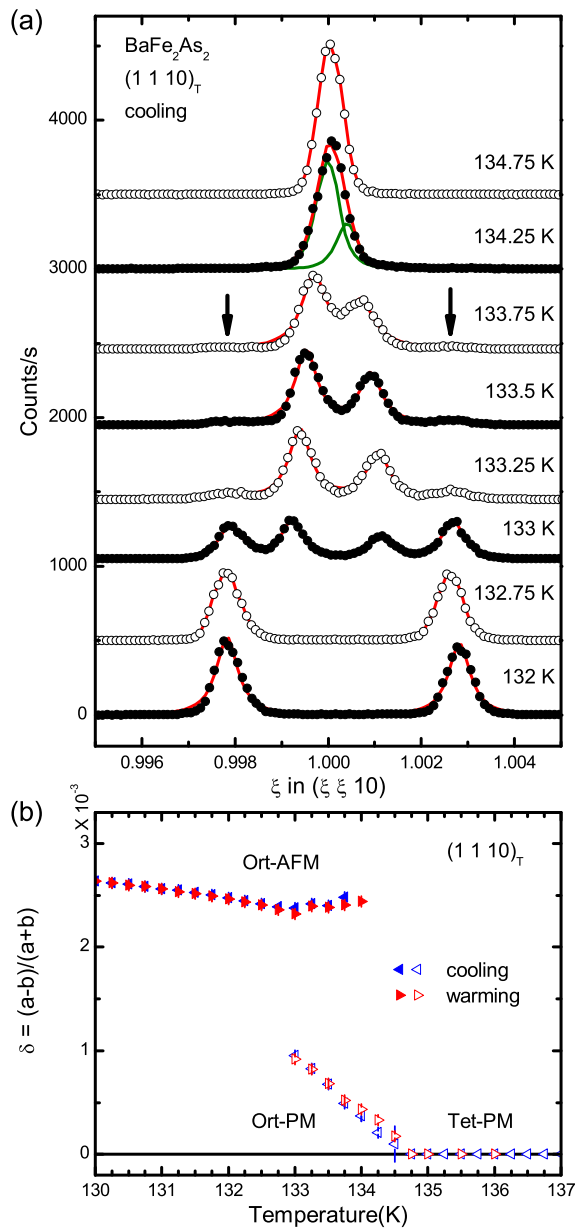}\\
\caption{(Color online) (a) X-ray diffraction scans, measured using
the laboratory source, along the [$\xi$ $\xi$ 0] direction through
the position of the tetragonal (1 1 10)$_T$ reflection for selected
temperatures in the parent BaFe$_2$As$_2$ upon cooling. The lines
present the fitted curves using a Lorentzian-squared line shape. The
two-component fit to broadened peaks is illustrated for $T$ = 134.25
K. The arrows denote the positions of peaks associated with Ort-AFM
as discussed in the text. At this temperature, the integrated
intensity of the Ort-AFM peaks are approximately 5$\%$ of the Ort-PM
diffraction peaks. (b) The orthorhombic distortion as a function of
temperature upon cooling and warming determined from fits to the (1
1 10) Bragg diffraction peak.} \label{parent}
\end{center}
\end{figure}

In Fig.~\ref{parent}(a), we display [$\xi$ $\xi$ 0] scans through
the (1 1 10)$_T$ charge peak, obtained using the laboratory source,
for the parent BaFe$_2$As$_2$ compound measured with temperature
steps of 0.25 K. Above the structural transition temperature, $T_S$
= 134.5 K, we observe a single sharp peak consistent with the
tetragonal structure. Upon cooling below $T_S$, the (1 1 10)$_T$
charge peak continuously broadens and, then, clearly splits at $T$ =
133.75 K concomitant with the abrupt appearance of two additional
peaks at this temperature [vertical arrows in Fig.~\ref{parent}(a)]
bracketing the two inner peaks. Upon further cooling, the splitting
of the two inner peaks evolves continuously as their intensities
decrease, whereas the positions of the outer peaks change only
slowly as their intensities increase. Below $T$ = 133.0 K, the two
inner peaks disappeared leaving only the outer peaks in evidence. We
note that these observations are qualitatively consistent with
similar diffraction measurements on an annealed sample of
BaFe$_2$As$_2$ recently reported by Rotundu
\emph{et~al}.\cite{Rotundu_2010} although the transition
temperatures for their annealed sample were approximately 5 K
higher.

Having described the temperature evolution of the diffraction peaks
qualitatively, it is useful at this point to introduce some labeling
of the corresponding phases. The high-temperature paramagnetic phase
is denoted as Tet-PM. Anticipating the results of our XRMS study, we
label the orthorhombic phase that evolves continuously over a very
narrow temperature range below $T_S$ [corresponding to the inner
pair of peaks in Fig.~\ref{parent}(a)] as Ort-PM. We further label
the orthorhombic phase that abruptly appears at $T$ = 133.75 K
[corresponding to the two outer bracketing peaks in
Fig.~\ref{parent}(a)] as Ort-AFM. Structurally, we assume that
Ort-PM and Ort-AFM differ only with respect to the values of their
lattice constants and orthorhombic distortion at a given
temperature.

Figure~\ref{parent}(b) describes the temperature evolution of these
phases. Upon cooling, a second-order transition from Tet-PM to
Ort-PM occurs at $T_S$ = 134.5 K followed by a first-order
transition to Ort-AFM at $T_N$ = 133.75 K. There is a region of
coexistence between Ort-AFM and Ort-PM from 133.75 K to 133.0 K, and
only the Ort-AFM phase is observed below this temperature. Upon
warming, Ort-PM appears at 133.0 K and coexists with Ort-AFM up to
$T'_N$ = 134.0 K, where Ort-AFM disappears. The orthorhombic
distortion associated with Ort-PM decreases continuously up to $T_S$
= 134.5 K, where Tet-PM is recovered. We find no hysteresis in the
transformations from Tet-PM to Ort-PM and $\lesssim$0.25 K
hysteresis associated with the appearance/disappearance of Ort-AFM.

\begin{figure}
\begin{center}
\includegraphics[clip, width=0.42\textwidth]{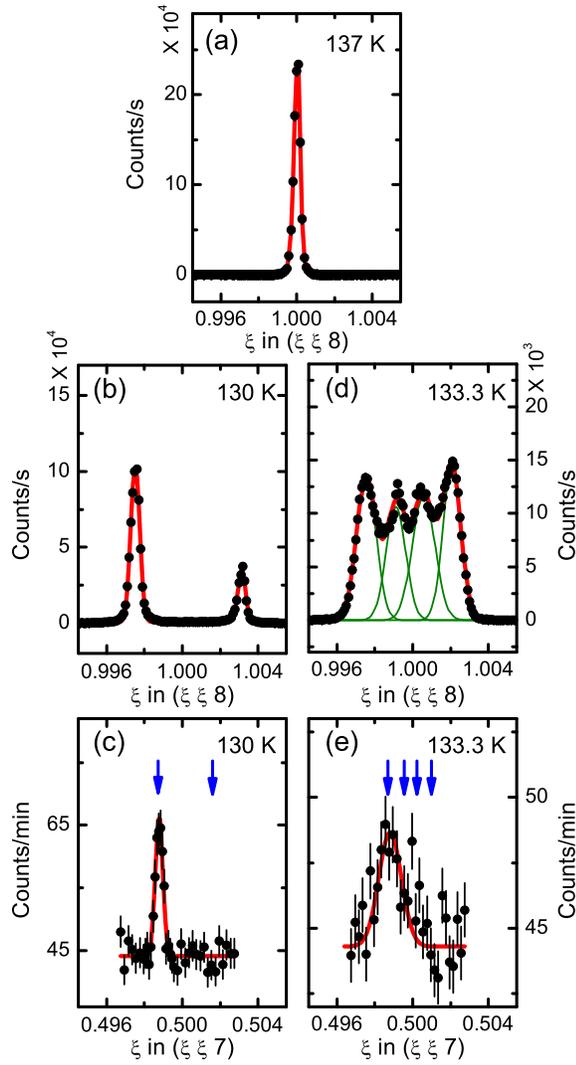}\\
\caption{(Color online) (a) The measured (1 1 8)$_T$ charge
diffraction peak above the structural/magnetic transitions. Panels
(b) and (c) show the (1 1 8)$_T$ charge peak and ($\frac{1}{2}$
$\frac{1}{2}$ 7)$_{T}$ magnetic peak at $T$ = 130 K, well below the
transition region. Panels (d) and (e) show the measured intensities
at the (1 1 8)$_T$ charge peak and ($\frac{1}{2}$ $\frac{1}{2}$
7)$_{T}$ magnetic positions at $T$ = 133.3 K. The arrows in (c) and
(e) indicate the calculated magnetic peak positions corresponding to
each of the charge peaks in (b) and (d), respectively.  The fitted
value for the width of the charge and magnetic peaks are the same.}
\label{magpeaks}
\end{center}
\end{figure}

In order to investigate the relationship between the structural
transition and AFM ordering in this system we have performed a
combined study using high-resolution x-ray diffraction and XRMS
measurements. These simultaneous measurements eliminate concerns
regarding disparities in the temperature calibration of sensors for
different experiments. Using the configuration at the APS described
in the last section we measured the scattering at both the charge
and magnetic Bragg peak positions for several temperatures close to
the structural transition. In Fig.~\ref{magpeaks}(a) we show a
[$\xi$ $\xi$ 0] scan through the (1 1 8)$_T$ charge Bragg peak at
$T$ = 137 K, well above the structural transition temperature. At
$T$ = 130 K, below $T_N$ and $T_S$, two well-separated peaks were
observed [Fig.~\ref{magpeaks}(b)]. These are the (2 0 8)$_O$ and (0
2 8)$_O$ charge Bragg peaks of the orthorhombic phase. The
difference in intensity arises from different populations of the
domains within the illuminated volume of the sample. At this same
temperature, a single peak is found at the (1 0 7)$_O$, magnetic
peak position for the orthorhombic phase, [Fig.~\ref{magpeaks}(c)]
in agreement with previous measurements of a magnetic propagations
vector given by \textbf{Q$_m$} = (1 0 1)$_O$ [($\frac{1}{2}$
$\frac{1}{2}$ 1)$_T$] with lattice constants $a_O$ $>$
$b_O$.~\cite{huang_neutron-diffraction_2008,wilson_neutron_2009,matan_anisotropic_2009,kofu_title_2009,kim_resonant_2010}
For simplicity, we will henceforth label all peaks with tetragonal
indices. Therefore, the (1 0 7)$_O$ magnetic Bragg peak will be
referred to as ($\frac{1}{2}$ $\frac{1}{2}$ 7)$_T$, keeping in mind
that the magnetic peaks are displaced from $\xi$ = $\frac{1}{2}$
because of the orthorhombic distortion.

The principal result conveyed in Figs.~\ref{magpeaks}(b) and (c) is
that the expected AFM order exists in the Ort-AFM phase. The
question, however, is whether this AFM order is also associated with
the Ort-PM intermediate phase. Figures~\ref{magpeaks}(d) and (e)
show [$\xi$ $\xi$ 0] scans through the (1 1 8)$_T$ charge and
($\frac{1}{2}$ $\frac{1}{2}$ 7)$_{T}$ magnetic peak positions at $T$
= 133.3 K. Similar to what was found in our laboratory-based
measurement [Fig.~\ref{parent}(a)] we observed four charge peaks
[two outer peaks from Ort-AFM and two inner peaks from Ort-PM].
However, Fig.~\ref{magpeaks}(e) displays only a single magnetic
peak.  The arrows in this panel denote the expected positions for
magnetic peaks associated with each of the charge peaks in
Fig.~\ref{magpeaks}(d) and we see that the magnetic peak is found at
a position that corresponds to one of the two outer peaks associated
with the Ort-AFM phase. This allows us to conclude that the magnetic
order is associated only with the Ort-AFM phase.  Taken together,
the x-ray diffraction and XRMS measurements suggest that: (1) The
orthorhombic distortion at $T_S$ is best described as a second-order
transition; (2) the structural and AFM transitions in the as-grown
BaFe$_2$As$_2$ compound are separated in temperature by
approximately 0.75 K and; (3) a first-order magnetic transition at
$T_N$ drives the discontinuity in the structural order parameter at
133.75 K.  To further substantiate these conclusions, we now turn to
a study of the evolution of the structural transition in
electron-doped BaFe$_2$As$_2$ compounds.

\subsection{High-resolution x-ray diffraction and resistance measurements of Ba(Fe$_{1-x}$Co$_{x}$)$_2$As$_2$
and Ba(Fe$_{1-x}$Rh$_{x}$)$_2$As$_2$}

\begin{figure}
\begin{center}
\includegraphics[clip, width=0.42\textwidth]{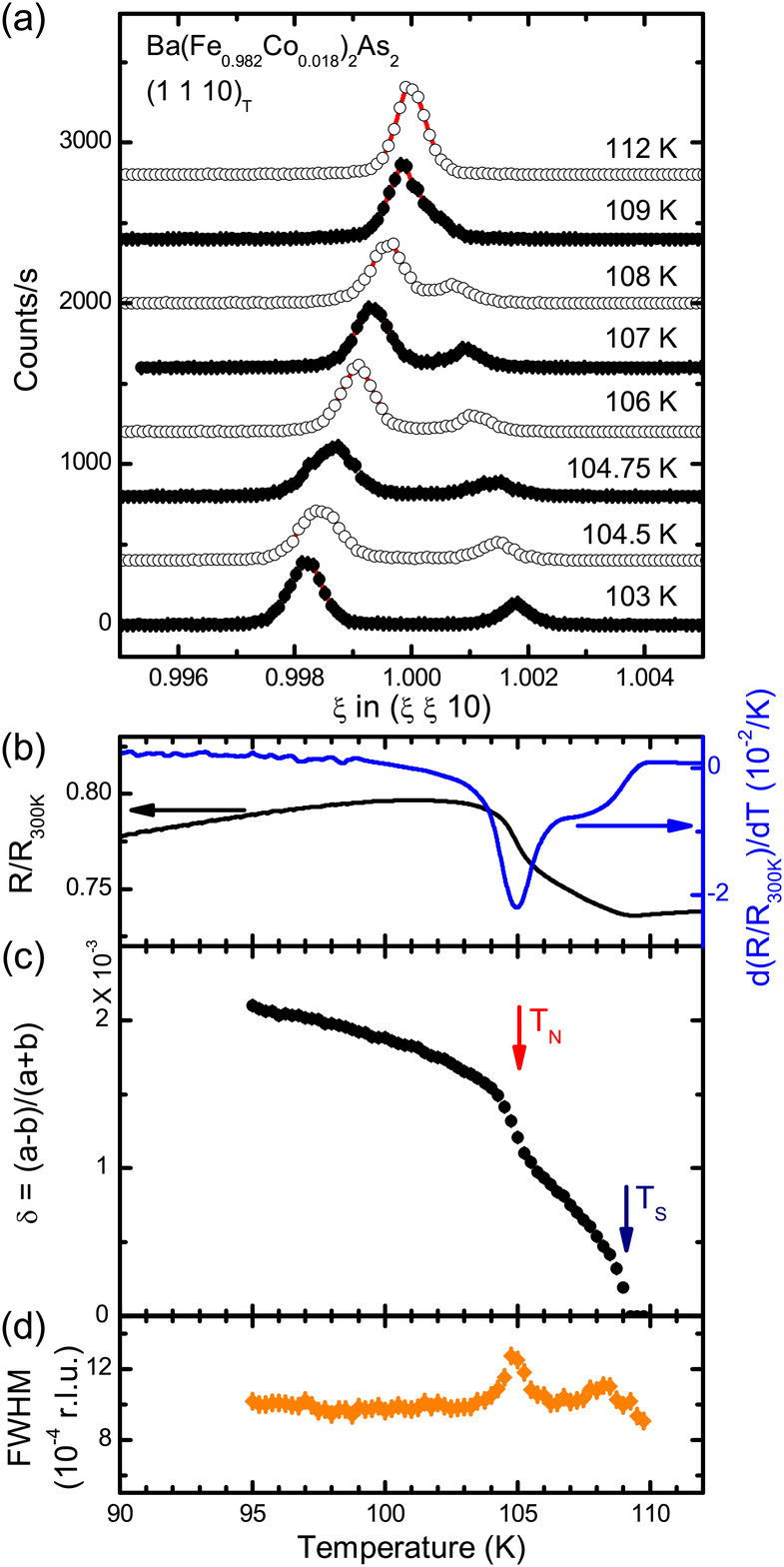}\\
\caption{(Color online) (a) X-ray data, (b) resistance (black line)
and its temperature derivative (blue line), (c) orthorhombic
distortion and (d) FWHM of the split (1 1 10)$_T$ Bragg peaks
measured for Ba(Fe$_{1-x}$Co$_x$)$_2$As$_2$ with $x$ = 0.018.  In
panel (c) the structural and magnetic transition temperatures are
marked.} \label{co_018}
\end{center}
\end{figure}

It has already been established that the substitution of Co or Rh
for Fe in the parent BaFe$_2$As$_2$ compound results in new and
interesting
behavior.\cite{canfield_feas_2010,ni_effects_2008,Chu_2009,Lester_2009,pratt_coexistence_2009,Christianson_2009,kreyssig_suppression_2010,nandi_anomalous_2010,fernandes_unconventional_2010}
As doping is increased, both the structural and magnetic transitions
are suppressed and split, with the structural transition occurring
at higher temperature.  In Ba(Fe$_{1-x}$Co$_{x}$)$_2$As$_2$, for Co
concentrations 0.03 $\leq$ $x$ $\leq$ 0.06, we enter a region of the
phase diagram where magnetism and superconductivity coexist and
compete.\cite{ni_effects_2008,fernandes_unconventional_2010}  Within
this region, the magnetic and structural transitions are
well-separated in temperature, and continuous in nature (see for
example, references ~\onlinecite{nandi_anomalous_2010} and
~\onlinecite{fernandes_unconventional_2010}).  It is, therefore,
interesting to probe the behavior of the structural and AFM
transitions, described above, as they evolve towards split
second-order transitions with doping.

\begin{figure}
\begin{center}
\includegraphics[clip, width=0.42\textwidth]{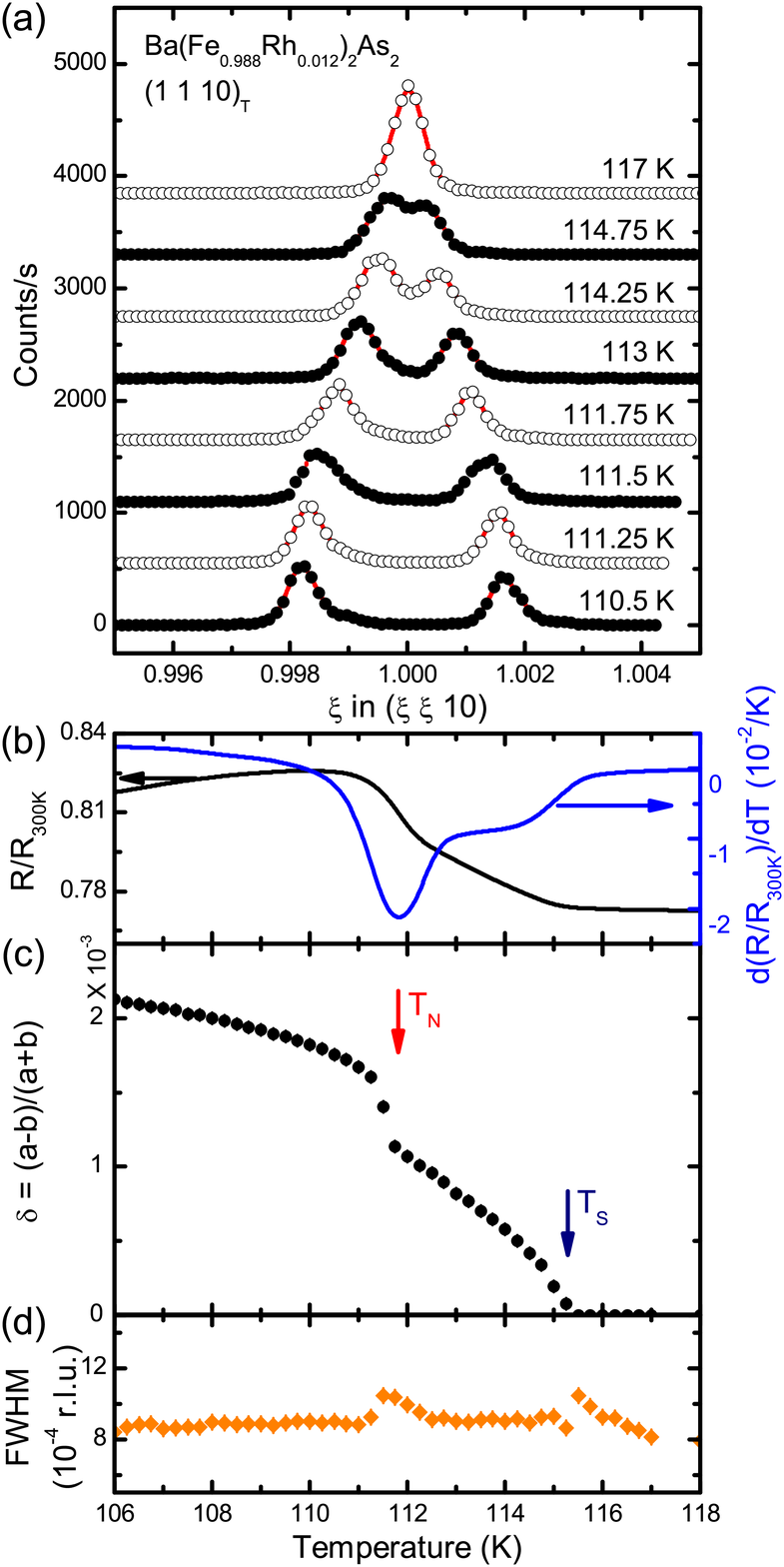}\\
\caption{(Color online) (a) X-ray data, (b) resistance (black line)
and its temperature derivative (blue line), (c) orthorhombic
distortion and (d) FWHM of the split (1 1 10)$_T$ Bragg peaks
measured for Ba(Fe$_{1-x}$Rh$_x$)$_2$As$_2$ with $x$ = 0.012.  In
panel (c) the structural and magnetic transition temperatures are
marked.} \label{rh_012}
\end{center}
\end{figure}

\begin{figure}
\begin{center}
\includegraphics[clip, width=0.42\textwidth]{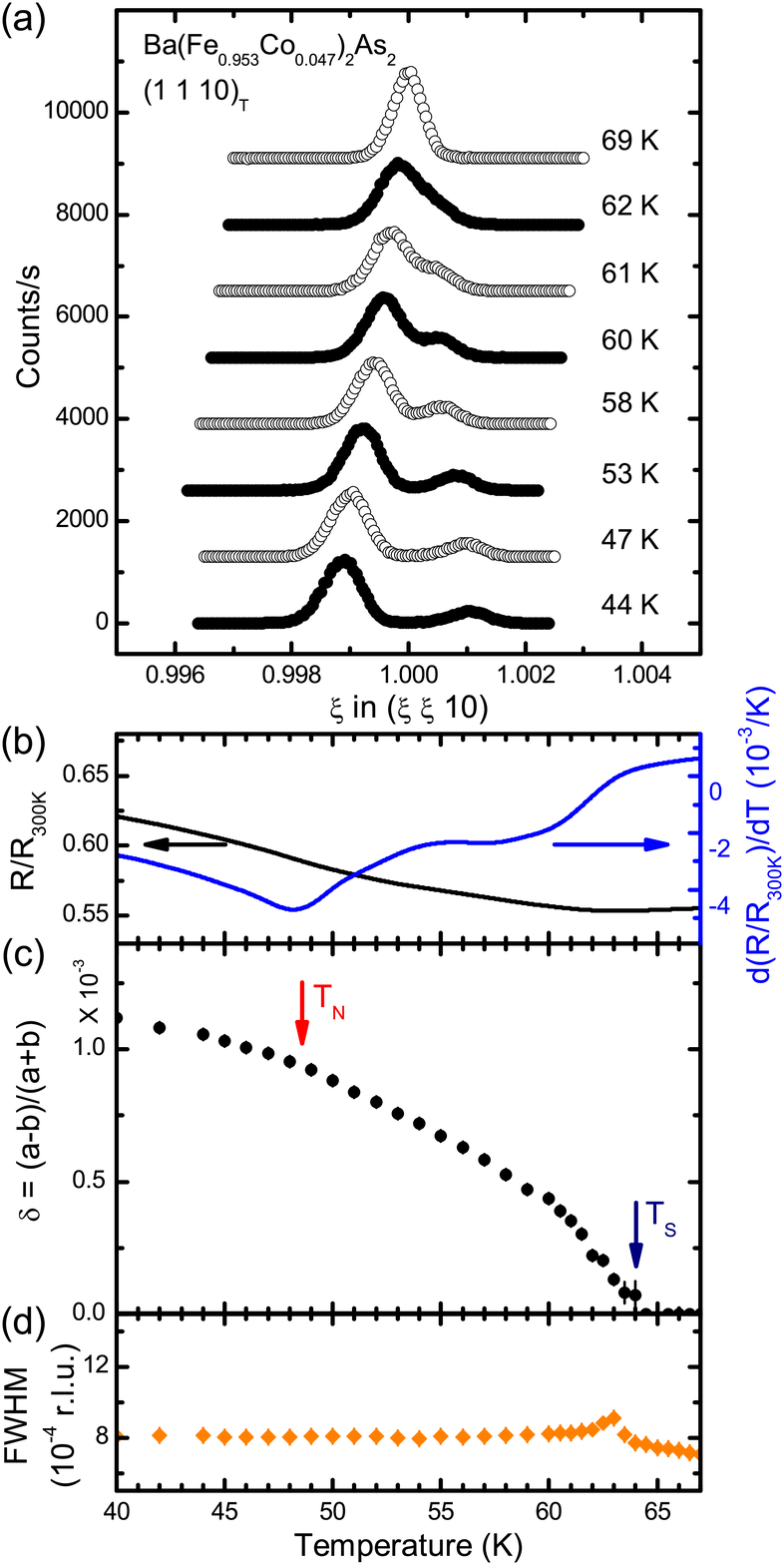}\\
\caption{(Color online) (a) X-ray data, (b) resistance (black line)
and its temperature derivative (blue line), (c) orthorhombic
distortion and (d) FWHM of the split (1 1 10)$_T$ Bragg peaks
measured for Ba(Fe$_{1-x}$Co$_x$)$_2$As$_2$ with $x$ = 0.047.  In
panel (c) the structural and magnetic transition temperatures are
marked.} \label{co_047}
\end{center}
\end{figure}

\begin{figure}
\begin{center}
\includegraphics[clip, width=0.42\textwidth]{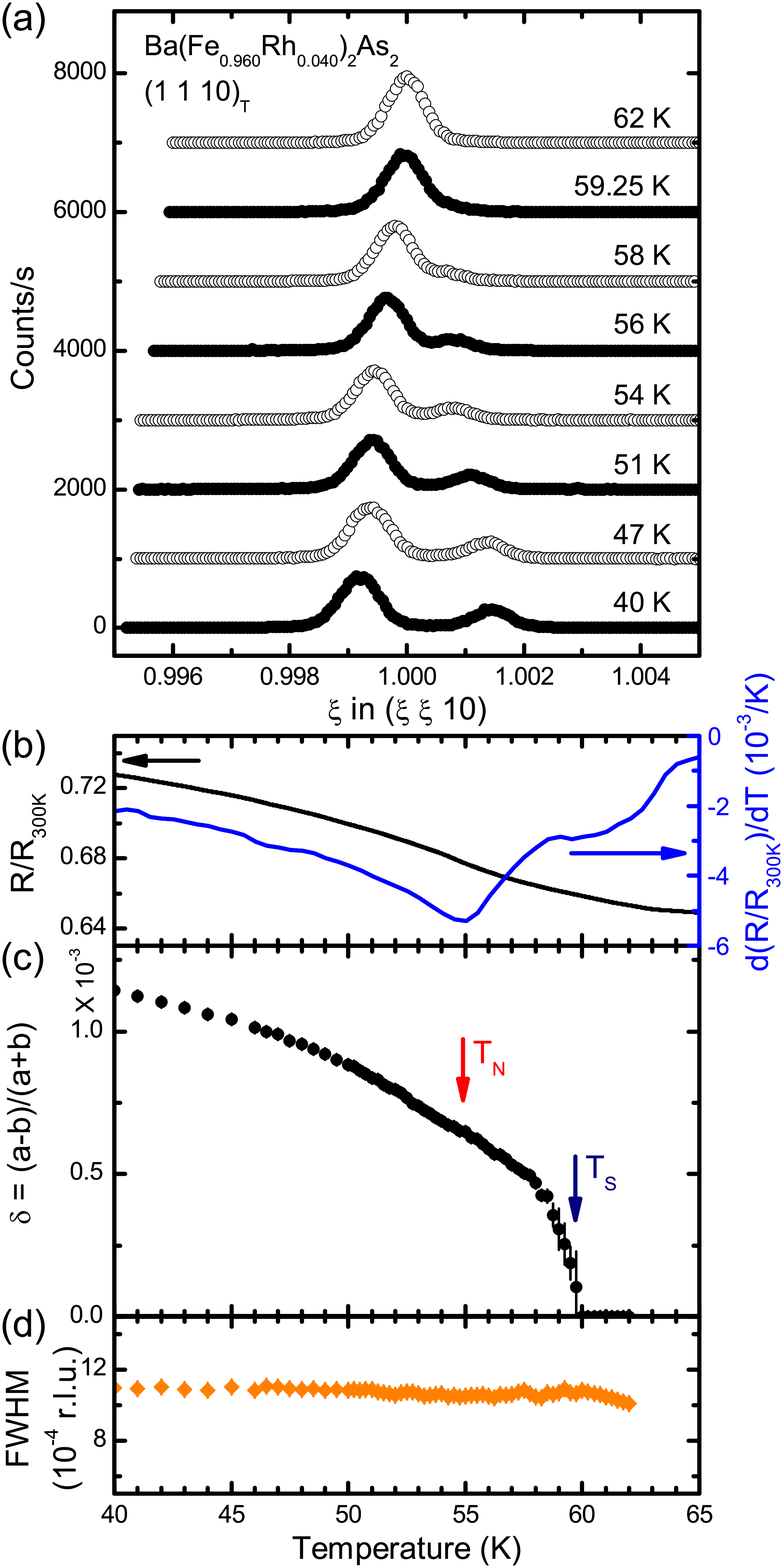}\\
\caption{(Color online) (a) X-ray data, (b) resistance (black line)
and its temperature derivative (blue line), (c) orthorhombic
distortion and (d) FWHM of the split (1 1 10)$_T$ Bragg peaks
measured for Ba(Fe$_{1-x}$Rh$_x$)$_2$As$_2$ with $x$ = 0.040.  In
panel (c) the structural and magnetic transition temperatures are
marked.} \label{rh_040}
\end{center}
\end{figure}

To this end, we have performed high-resolution x-ray diffraction
measurements on four doped BaFe$_2$As$_2$ samples:
Ba(Fe$_{1-x}$Co$_{x}$)$_2$As$_2$ for $x$ = 0.018 and 0.047, and
Ba(Fe$_{1-x}$Rh$_{x}$)$_2$As$_2$ for $x$ = 0.012 and 0.040.
Figures~\ref{co_018} through \ref{rh_040} display the raw
diffraction data, the orthorhombic distortion ($\delta$) and
diffraction peak widths derived from fits to the data, and the
electrical resistance measured as a function of temperature.

Turning first to the compounds at lower doping concentrations,
Ba(Fe$_{0.982}$Co$_{0.018}$)$_2$As$_2$ and
Ba(Fe$_{0.988}$Rh$_{0.012}$)$_2$As$_2$ (Figs.~\ref{co_018} and
\ref{rh_012}), below $T_S$ both samples manifest a lattice
distortion that evolves continuously as temperature is lowered,
until the onset of magnetic ordering where a step-like feature in
the structural order parameter ($\delta$) is observed. At $T_N$ a
distinct broadening of the split orthorhombic diffraction peaks is
evident over a narrow range in temperature. In contrast, the
temperature dependence of the order parameter and peak widths for
the higher doping concentrations,
Ba(Fe$_{0.953}$Co$_{0.047}$)$_2$As$_2$ and
Ba(Fe$_{0.960}$Rh$_{0.040}$)$_2$As$_2$, is decidedly different near
$T_N$ (Figs.~\ref{co_047} and \ref{rh_040}).  For these samples, the
structural distortion evolves continuously, with only a mild kink in
evidence at $T_N$ and without the attendant peak broadening at
$T_N$.

\begin{figure}
\begin{center}
\includegraphics[clip, width=0.42\textwidth]{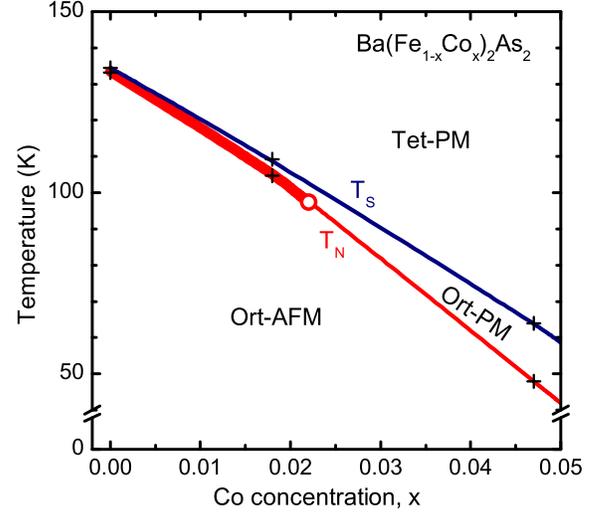}\\
\caption{(Color online) Diagram showing the nature of the structural
and magnetic phase transitions for Ba(Fe$_{1-x}$Co$_x$)$_2$As$_2$ at
$T_S$ and $T_N$, respectively. The thick line denotes a first-order
transition whereas the thinner lines represent second-order
transitions. The crosses denote values for $T_S$ and $T_N$
determined from our measurements. The open circle denotes the
approximate position of a tricritical point as described in the
text.} \label{PD}
\end{center}
  \end{figure}

The differences between the lower and higher doping concentrations
are consistent with a change in the nature of the magnetic
transition from first-order for low doping, to second-order for
higher doping levels.  At low-doping, as for the parent
BaFe$_2$As$_2$ compound, there is a second-order transition from the
tetragonal phase to the Ort-PM structure as temperature is decreased
below $T_S$. The step in the orthorhombic distortion ($\delta$) at
$T_N$ is a consequence of the abrupt appearance of Ort-AFM
coincident with a first-order AFM transition. We note that
throughout this temperature range only two broadened peaks are
observed in contrast to what was shown above for the parent
compound.  This is expected, however, since the larger separation of
$T_S$ and $T_N$ allows $\delta$ to evolve to a value for the Ort-PM
phase that is close to its magnitude for the Ort-AFM phase. The
anomalous increase in the widths of the x-ray diffraction peaks at
$T_N$ arises from the coexistence and near coincidence in position
of the Ort-AFM and Ort-PM diffraction peaks over a narrow
temperature range.  For the higher doping levels, within our
experimental resolution, the absence of a distinct step in $\delta$
or peak broadening at $T_N$ is consistent with second-order AFM and
structural transitions as found for electron-doped BaFe$_2$As$_2$
previously.\cite{fernandes_unconventional_2010,nandi_anomalous_2010}

We summarize our results in Fig. \ref{PD} which displays a phase
diagram for Ba(Fe$_{1-x}$Co$_{x}$)$_2$As$_2$ focusing on the
concentration range of the present study.  The structural
transition, over the entire range is second-order, whereas the
magnetic transition changes from first-order to second-order at a
tricritical point as discussed below.

\section{Discussion}
\begin{figure}
\begin{centering}
\includegraphics[width=0.9\columnwidth]{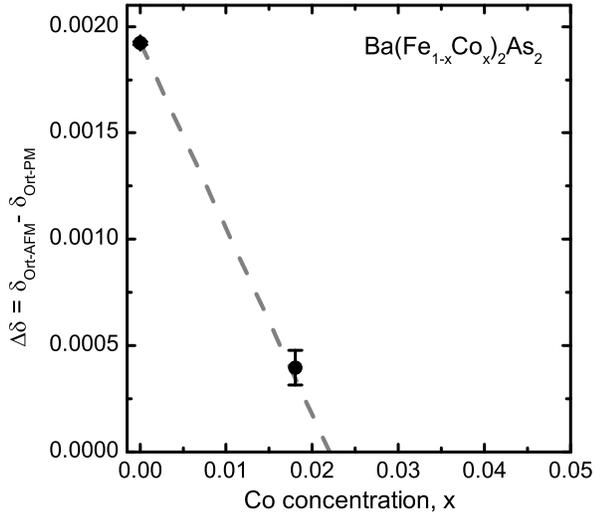}
\par\end{centering}
\caption{Jump of the orthorhombic order parameter
$\Delta\delta\equiv\delta_{\mathrm{Ort-AFM}}-\delta_{\mathrm{Ort-PM}}$
across the first-order magnetic transition, as function of $x$.
The linear relationship $\Delta\delta\propto\left(x-x_{\mathrm{tri}}\right)$
(dashed line) follows from the mean-field solution of Eq.
\ref{F_eff}. \label{fig_tricritical}}
\end{figure}

To understand the existence, and estimate the location, of a
magnetic tricritical point in the phase diagram of
$\mathrm{Ba\left(Fe_{1-x}Co_{x}\right)_{2}As_{2}}$, we can, at
first, rationalize the interplay between the magnetic and elastic
degrees of freedom in terms of a simple Ginzburg-Landau model,
similar to what was done by Cano \emph{et al.} in Ref.
~\onlinecite{Cano_10}. We start from the effective free energy:

\begin{equation}
F_{\mathrm{eff}}=\left(\frac{a_{\delta}}{2}\delta^{2}+\frac{u_{\delta}}{4}\delta^{4}\right)+\left(\frac{a_{m}}{2}m^{2}+\frac{u_{m}}{4}m^{4}\right)-\lambda\delta\:
m^{2}\label{F_eff}\end{equation} with
$a_{\delta}=a_{\delta,0}\left(T-T_{S,0}\right)$,
$a_{m}=a_{m,0}\left(T-T_{N,0}\right)$, and positive constants
$u_{\delta}$, $u_{m}$, $\lambda$. Here, $m$ is the antiferromagnetic
order parameter and $T_{S,0}$, $T_{N,0}$ denote the bare structural
and magnetic transition temperatures respectively. For $T_{S,0} <
 T_{N,0}$, this model describes a simultaneous magnetic/structural
first-order transition. However, for $T_{S,0} > T_{N,0}$, this model
describes a second-order structural transition at $T_{S} = T_{S,0}$,
followed by a magnetic transition at $T_{N}$
($T_{N,0}<T_{N}<T_{S}$), which can be either first-order or
second-order. Considering that $T_{N}$ and $T_{S}$ change with
doping, the magnetic tricritical point takes place at the doping
concentration $x_{\mathrm{tri}}$ where
$u_{m}a_{\delta,0}\left(T_{S}-T_{N}\right)=\lambda^{2}$.
Experimentally, we know that $\left(T_{S}-T_{N}\right)$ increases
with doping, $x$. Therefore, it is straightforward to conclude from
the mean-field solution of Eq. (\ref{F_eff}) that, close to the
magnetic tricritical point, the jump in the orthorhombic order
parameter across the first-order magnetic transition changes with
doping as
$\Delta\delta\equiv\delta_{\mathrm{Ort-AFM}}-\delta_{\mathrm{Ort-PM}}\propto\left(x-x_{\mathrm{tri}}\right)$.
Extrapolating this linear relation and using the values of
$\Delta\delta$ from Figs. \ref{parent} and \ref{co_018}, we estimate
that the magnetic tricritical point is located at
$x_{\mathrm{tri}}\approx0.022$, as shown in Fig.
\ref{fig_tricritical}.

\begin{figure}
\begin{centering}
\includegraphics[width=0.9\columnwidth]{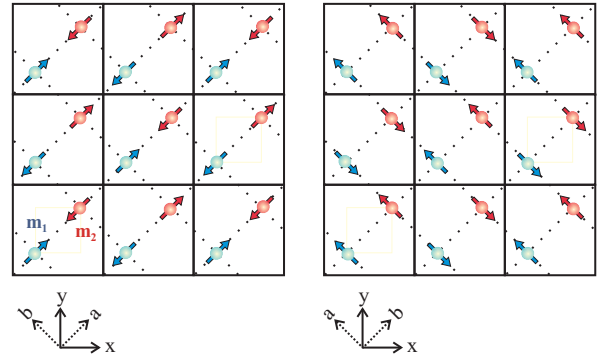}
\par\end{centering}
\caption{(Color online) The two magnetic ground states of the iron
pnictides, characterized by stripes along two orthogonal directions,
expressed in terms of two interpenetrating AFM sublattices with
N\'{e}el vectors $\mathbf{m}_{1}$ and $\mathbf{m}_{2}$. Notice that
the $\left(x,y\right)$ coordinate system used here refers to the
unit cell with two Fe atoms and is rotated by $45^{\circ}$ with
respect to the $\left(a,b\right)$ coordinate system relative to the
single-Fe atom unit cell. \label{fig_AFM_sublattices}}
\end{figure}

The main issue with the model in Eq. (\ref{F_eff}), however, is that
it requires a fine tuning of the independent structural and magnetic
transition temperatures $T_{S,0}$ and $T_{N,0}$ across the phase
diagram. In all of the phase diagrams of the iron pnictide
compounds, it is observed that the two transition lines track each
other very closely, even within the superconducting dome.
\cite{Fernandes_2010_2,nandi_anomalous_2010} This suggests that
these two states are strongly coupled, rather than independent, as
assumed by the previous model. To address this issue, it has been
proposed that the particular magnetic structure of the iron
pnictides gives rise to emergent Ising-nematic degrees of freedom
that couple to the lattice, inducing the tetragonal-to-orthorhombic
phase transition \cite{Fang_08,Xu_08,Fernandes_2010_2}. In the
magnetically ordered phase, there are two degenerate ground states
characterized by in-plane spin stripes along each of the two
orthogonal Fe-Fe bond directions. These ground states can be
described in terms of two interpenetrating AFM sublattices with
staggered magnetization $\mathbf{m}_{1}$ and $\mathbf{m}_{2}$, such
that $\mathbf{m}_{1}$ is either parallel or antiparallel to
$\mathbf{m}_{2}$ (see Fig. \ref{fig_AFM_sublattices}).

Within this description, the magnetic free energy of the system
$F_{\mathrm{mag}}$ has contributions from each sublattice $F_{i}$ and from
the coupling between them, $F_{12}$. The former is given by:

\begin{equation}
F_{i}=\frac{1}{2}\int\frac{d^{3}q}{\left(2\pi\right)^{3}}\:\chi_{i}^{-1}\left(\mathbf{q}\right)\mathbf{m}_{i}\left(\mathbf{q}\right)\cdot\mathbf{m}_{i}\left(-\mathbf{q}\right)+\frac{u}{4}\int\frac{d^{3}x}{v}\:\mathbf{m}_{i}^{4}\label{Fi}\end{equation}
where
$\chi_{i}\left(\mathbf{q}\right)=\chi_{0}\left(r_{0}+q_{\parallel}^{2}a^{2}-2\eta_{z}\cos
q_{\perp}c\right)^{-1}$ is the static susceptibility of each sublattice,\cite{Fernandes_2010_2}
$u$ is a positive coupling constant, and the momentum $\mathbf{q}$
is measured relative to the magnetic ordering vector. Here,
$\chi_{0}^{-1}$ is a magnetic energy scale, $r_{0}$ measures the
distance to the magnetic critical point, $a$ and $c$ are the lattice
parameters of the unit cell of volume $v$ containing two Fe atoms, and
$\eta_{z}$ is the inter-plane AFM coupling. The coupling between the
two sublattices is given by:

\begin{equation}
F_{12}=-\frac{g}{2}\int\frac{d^{3}x}{v}\left(\mathbf{m}_{1}\cdot\mathbf{m}_{2}\right)^{2}\label{F12}\end{equation}
 with $g>0$, favoring configurations where $\mathbf{m}_{1}$ and
$\mathbf{m}_{2}$ are either parallel or antiparallel. In a
description of the magnetically ordered phase in terms of localized
moments, this term originates from quantum and thermal spin
fluctuations \cite{Chandra_90}. On the other hand, within an
itinerant approach, where the magnetic moments arise from the
conduction electrons, the same term appears as a consequence of the
ellipticity of the electron pockets \cite{EandC_2010}.

The coupling (\ref{F12}) between the sublattices gives rise to an
emergent Ising-nematic degree of freedom
$\varphi=\mathbf{m}_{1}\cdot\mathbf{m}_{2}$ \cite{Chandra_90}, which
may be finite even in the absence of magnetic order (i.e.
$\left\langle \varphi\right\rangle $ $\neq$ 0, but $\left\langle
\mathbf{m}_{i}\right\rangle =0$) as long as the magnetic
fluctuations are strong enough \cite{Fernandes_2010_2}. A finite
value, $\left\langle \mathbf{m}_{1}\cdot\mathbf{m}_{2}\right\rangle
\neq0$, breaks the Ising symmetry embedded in Eq. (\ref{F12}) and,
consequently, the tetragonal symmetry. This can
be seen explicitly through the magneto-elastic term:

\begin{equation}
F_{\mathrm{mag-el}}=\lambda\int\frac{d^{3}x}{v}\:\delta\left(\mathbf{m}_{1}\cdot\mathbf{m}_{2}\right)\label{F_mag_el}\end{equation}
where $\lambda>0$ is the magneto-elastic coupling and $\delta$ is
the orthorhombic distortion. From the bilinear coupling of $\delta$
and $\varphi$ in Eq. (\ref{F_mag_el}), both the nematic and structural
transitions are simultaneous. This mechanism for the tetragonal-to-orthorhombic
transition explains why the magnetic and structural
transitions track each other closely in all the phase diagrams of
the iron pnictides. Furthermore, it also explains several
experimental observations, such as the softening of the lattice in
the tetragonal phase and its hardening in the superconducting
state,\cite{Fernandes_2010_2} as well as the suppression of the
orthorhombic distortion below the superconducting transition
temperature.\cite{nandi_anomalous_2010}

In the case where the elastic free energy is harmonic,
$F_{el}=C_{s}\delta^{2}/2$ [where $C_{s}$ is the shear modulus] the
only effect of the elastic degree of freedom is to renormalize the
sublattice coupling constant $g$ in Eq. (\ref{F12}), yielding
$g\rightarrow g+\lambda^{2}/C_{s}$ \cite{Fernandes_2010_2}. In a
mean-field approach, this implies that the two transitions remain
split and second order. Although fluctuations could induce a
simultaneous first-order transition \cite{Gorkov_09,Qi_09}, it is
unclear whether they could explain a second-order structural
transition split from a first-order magnetic transition, as our data
for low doping levels in $\mathrm{BaFe_{2}As_{2}}$ suggests.

To account for our observations, we move beyond the harmonic lattice
approximation to consider the effects of anharmonic elastic terms (for
more details on the formalism, see Ref. ~\onlinecite{Fadda_02}).
In the tetragonal phase, the most general form of the free energy
can be written as
$F_{\mathrm{el}}=\frac{1}{2}\bar{C}_{ij}\epsilon_{i}\epsilon_{j}+\frac{1}{6}\bar{C}_{ijk}\epsilon_{i}\epsilon_{j}\epsilon_{k}$,
where $\bar{C}_{ij}$ are given in terms of the six independent
elastic constants and the strain components $\epsilon_{i}$ are:

\begin{eqnarray}
\epsilon_{1}= & u_{xx}+u_{yy}+u_{zz} & \,;\quad\quad\epsilon_{4}=2u_{yz}\nonumber \\
\epsilon_{2}= & \left(u_{xx}+u_{yy}-2u_{zz}\right)/6 & \,;\quad\quad\epsilon_{5}=2u_{xz}\nonumber \\
\epsilon_{3}= & \left(u_{xx}-u_{yy}\right)/\sqrt{2} &
\,;\quad\quad\epsilon_{6}=2u_{xy}\label{strain_tensor}\end{eqnarray}
with $u_{ij}=\left(\partial_{i}u_{j}+\partial_{j}u_{i}\right)/2$ and
$\mathbf{u}=\left(u_{x},u_{y},u_{z}\right)$ denoting the
displacement vector. In this notation, we have the orthorhombic
distortion $\delta=\epsilon_{6}/2$ and the shear modulus
$C_{s}\equiv4\bar{C}_{66}$. Since we are interested in describing
the transition from the tetragonal to the orthorhombic phase, we
retain only the essential anharmonic terms that contain
$\epsilon_{6}$:

\begin{eqnarray}
F_{\mathrm{el}} & = & \frac{1}{2}\bar{C}_{11}\epsilon_{1}^{2}+\frac{1}{2}\bar{C}_{22}\epsilon_{2}^{2}+\bar{C}_{12}\epsilon_{1}\epsilon_{2}+\frac{1}{2}\bar{C}_{44}\left(\epsilon_{4}^{2}+\epsilon_{5}^{2}\right)+\nonumber \\
 &  & \frac{1}{2}\bar{C}_{66}\epsilon_{6}^{2}+\frac{1}{2}\left[\bar{C}_{166}\epsilon_{1}+\bar{C}_{266}\epsilon_{2}\right]\epsilon_{6}^{2}+\bar{C}_{456}\epsilon_{4}\epsilon_{5}\epsilon_{6}\label{F_el}\end{eqnarray}
 Minimization with respect to the other strain components yields an
effective elastic free energy in terms only of
$\epsilon_{6}=2\delta$:

\begin{equation}
F_{\mathrm{el}}\left[\delta\right]=\frac{1}{2}C_{s}\delta^{2}+\frac{1}{4}U\delta^{4}+\frac{1}{6}W\delta^{6}\label{F_el_eff}\end{equation}
 where we included the sixth-order term $W>0$ to ensure stability
of the functional. Note that since:\cite{Fadda_02}

\begin{equation}
\frac{U}{16} = U_{0} - \frac{\left( \bar{C}_{22}\bar{C}_{166}^{2} - 2\bar{C}_{12}\bar{C}_{166}\bar{C}_{266} + \bar{C}_{11}\bar{C}_{266}^{2} \right)}{2\left( \bar{C}_{11} \bar{C}_{22} - \bar{C}_{12}^{2} \right)} \label{quart_F_el_eff}\end{equation}
the fourth-order coefficient can be
negative, depending on the magnitudes of the anharmonic terms
$\bar{C}_{ijk}$. Here, $U_0$ is the bare coefficient coming from higher-order quartic anharmonic terms.

In what follows, we consider all elastic coefficients to be
temperature independent, and that the only minimum of Eq.
(\ref{F_el_eff}) is at $\delta=0$. Thus, different from the model in
Eq. (\ref{F_eff}), the system has no intrinsic structural instability,
and the elastic phase transition results solely from the
magneto-elastic coupling in Eq. (\ref{F_mag_el}). In the case of a
harmonic lattice, it was shown that nematic fluctuations renormalize
the shear modulus in the tetragonal phase, making it vanish when the
magnetic correlation length achieves a threshold value
\cite{Fernandes_2010_2}. Here, not only $C_{s}$ will be renormalized
by nematic fluctuations, but also the anharmonic term $U$ in Eq.
(\ref{F_el_eff}), which gives rise to a much richer phase diagram.

To calculate the total free energy of the system, we use the
'mean-field $1/N$ approach' discussed elsewhere
\cite{Fang_08,Fernandes_2010_2}. Basically, we assume that the
magnetic order parameter $\mathbf{m}_{i}$ has $N$ components and
expand to leading order for large $N$. We then obtain
self-consistent equations involving the magnetic correlation length
$\xi$, the nematic order parameter $\varphi$, the magnetic order
parameter
$m=\left|\langle\mathbf{m}_{1}\rangle\right|=\left|\langle\mathbf{m}_{2}\rangle\right|$,
and the orthorhombic distortion $\delta$. The latter is obtained by
minimizing the effective elastic free energy
$F_{\mathrm{eff}}=F_{\mathrm{el}}+\tilde{F}$, where $\tilde{F}$ is
an implicit function of $\delta$, arising from the $1/N$ solution of
the magnetic problem with free energy
$F_{\mathrm{mag}}+F_{\mathrm{mag-el}}$. Thus, $\tilde{F}$ describes
how magnetism changes the elastic free energy.

\begin{figure}
\begin{centering}
\includegraphics[width=0.9\columnwidth]{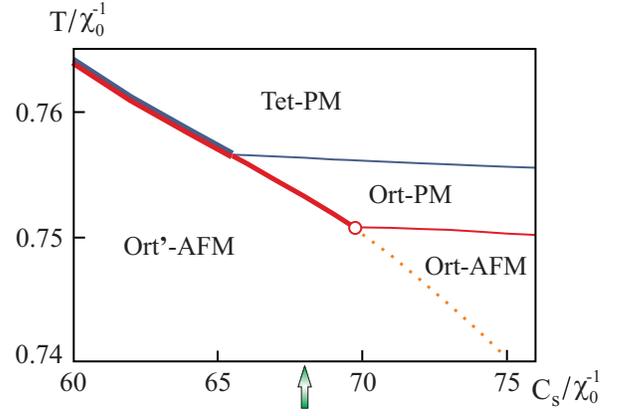}
\par\end{centering}
\caption{(Color online) Phase diagram of the system with anharmonic
elastic terms. $T$ denotes temperature, $C_{s}$ is the bare shear
modulus, and $\chi_{0}^{-1}$ is a magnetic energy scale. Thin
(thick) lines refer to second-order (first-order) phase transitions,
with the red (blue) lines denoting magnetic (structural)
transitions; the simultaneous first-order transition line is denoted
by the double line. We use the notation Ort' to emphasize that the
orthorhombic distortion jumps across the first-order magnetic
transition. The orange dotted line signals the occurrence of a jump
in both the magnetic and orthorhombic order parameters, without
symmetry breaking. The open circle refers to the magnetic
tricritical point, while the arrow indicates the value of $C_{s}$
for which we calculate the temperature dependence of both the
magnetic and orthorhombic order parameter (see
Fig.~\ref{fig_order_parameters}). \label{fig_phase_diagram}}

\end{figure}

We choose parameters that yield relative temperatures and jumps
comparable to those observed experimentally (see Section
IIIB).\cite{parameters} In particular, we take $U<0$ in Eq.
(\ref{F_el_eff}); once again, we stress that the values used for $U$
do not introduce any local minimum other than $\delta=0$ in the bare
$F_{el}$. To illustrate the richness of the resulting phase diagram,
in Fig.~\ref{fig_phase_diagram} we show the results obtained after
fixing all parameters but the bare shear modulus $C_{s}$. For
smaller values of $C_{s}$, the system undergoes a simultaneous
first-order structural/magnetic transition from the
tetragonal/paramagnetic phase to the orthorhombic/antiferromagnetic
phase. This corresponds to a direct first-order transition from the
Tet-PM phase to the Ort-AFM phase, which has not been observed in
the experiments reported here.

\begin{figure}
\begin{centering}
\includegraphics[width=0.9\columnwidth]{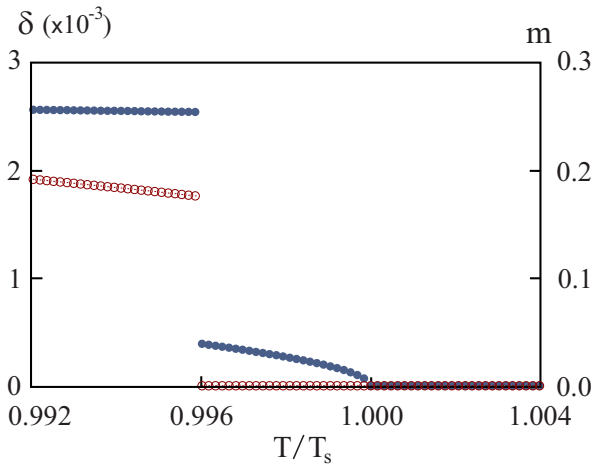}
\par\end{centering}
\caption{(Color online) Magnetic ($m$, open symbols) and
orthorhombic ($\delta$, filled symbols) order parameters as function
of temperature $T$ (in units of the structural transition
temperature $T_{s}$) for the system indicated by the green arrow in
the phase diagram of Fig.~\ref{fig_phase_diagram}.
\label{fig_order_parameters}}
\end{figure}

As the bare shear modulus increases, the two transitions split: at
higher temperatures, the system undergoes a \emph{second-order}
structural transition and then a \emph{first-order} magnetic
transition at lower temperature. The latter is accompanied by a
discontinuity in the orthorhombic order parameter $\delta$ due to
the magneto-elastic coupling. This is precisely the sequence
observed in our experiments described in Section III for the parent
BaFe$_{2}$As$_{2}$ and doped compounds for low doping concentrations
(Tet-PM $\rightarrow$ Ort-PM $\rightarrow$ Ort-AFM). Note that this
is not another structural phase transition, but a consequence of the
first-order character of the magnetic transition. To show this
explicitly, in Fig.~\ref{fig_order_parameters}, we plot both
$\delta$ and $m$ as function of temperature for the value of $C_{s}$
indicated by the arrow in Fig.~\ref{fig_phase_diagram}. Not only is
the relative size of the step comparable to that measured
experimentally for BaFe$_2$As$_2$, but also the relative temperature
at which the step occurs (see Fig.~\ref{parent}(b), where
$T_{s}\approx134.6$ K and $T_{N}\approx134$ K). The discontinuity in
the orthorhombic distortion accompanying the first-order magnetic
transition is a very general feature that holds regardless of the
specific values of the parameters. Thus, it supports our
interpretation that the experimental data in Fig.~\ref{parent} on
the parent compound, BaFe$_{2}$As$_{2}$, describe a second-order
structural transition followed by a first-order magnetic transition.

Returning to the phase diagram of Fig.~\ref{fig_phase_diagram}, we
note that as the shear modulus is increased even further, the
transitions remain split but the magnetic transition becomes
second-order, as it is observed for higher doping concentrations in
Ba(Fe$_{1-x}$Co$_{x}$)$_{2}$As$_{2}$ and
Ba(Fe$_{1-x}$Rh$_{x}$)$_{2}$As$_{2}$. At low temperatures, there is
another line that marks a simultaneous jump in both the magnetic and
the orthorhombic order parameter, without any symmetry breaking.
However, we point out that in our $1/N$ approach we have not taken
into account a key feature of the magnetically ordered state: the
reconstruction of the Fermi surface. For instance, we note that
$x_{\mathrm{tri}} \approx 0.02$ is close to the composition where
evidence for a Lifshitz transition, below $T_N$, in
Ba(Fe$_{1-x}$Co$_{x}$)$_2$As$_2$ has been reported by thermoelectric
power, Hall coefficient measurements and angle-resolved
photoemission.\cite{Mun_2009,Changliu_2010} Therefore, features in
our model deep in the magnetically ordered phase, such as this
extended line, are likely to change once the reconstruction of the
Fermi surface is considered. For instance, one possibility is that
this extended line terminates at a finite temperature critical
point.

Furthermore, although in the phase diagram of Fig.~\ref{fig_phase_diagram}
we changed only the bare shear modulus, it
is unlikely that this is the only modified parameter as doping is
introduced in the parent compound. In particular, the increase in
the splitting between the transitions is much more modest in Fig.~\ref{fig_phase_diagram}
than found experimentally (Fig.~\ref{PD}).
It is possible, then, that other parameters controlling the
splitting are also changed, such as the magneto-elastic coupling,
$\lambda$, and the inter-plane magnetic coupling, $\eta_{z}$. The
main objective of the phase diagram presented here is to illustrate
the various possible phase transitions once anharmonic elastic terms
are taken into account. It is interesting to note that, in our
simple phase diagram, systems with softer lattices are more likely
to display simultaneous first-order transitions. Indeed,
CaFe$_{2}$As$_{2}$, which is significantly softer than
BaFe$_{2}$As$_{2}$, presents relatively strong simultaneous
first-order transitions.\cite{goldman_lattice_2008}

\section{Summary}
Our high-resolution x-ray diffraction and XRMS studies of
BaFe$_2$As$_2$ have provided several new insights concerning the
nature of the structural and magnetic transitions in the fascinating
\emph{122} iron pnictide family.  First, we find that the
orthorhombic distortion at $T_S$ is best described as a second-order
transition and that the structural and AFM transitions in the
as-grown BaFe$_2$As$_2$ compound are separated in temperature by
approximately 0.75 K.  We propose that a first-order magnetic
transition at $T_N$ drives the discontinuity in the structural order
parameter at 133.75 K, and this is consistent with our measurements
of the evolution  of the character of the transitions in Co- and
Rh-doped BaFe$_2$As$_2$.  Using these results, we can provide an
estimate of the position of a tricritical point in the phase diagram
of Ba(Fe$_{1-x}$Co$_{x}$)$_{2}$As$_{2}$.  Finally, we employ a
mean-field approach to show that our measurements can be explained
by the inclusion of an anharmonic term in the elastic free energy
and magneto-elastic coupling in the form of an emergent
Ising-nematic degree of freedom.

We acknowledge valuable discussions with J. Lang. This work was
supported by the Division of Materials Sciences and Engineering,
Office of Basic Energy Sciences, U.S. Department of Energy. Ames
Laboratory is operated for the U.S. Department of Energy by Iowa
State University under Contract No. DE-AC02-07CH11358. Use of the
Advanced Photon Source was supported by the US DOE under Contract
No. DE-AC02-06CH11357.

\bibliographystyle{apsrev}
\bibliography{split}

\begin{thebibliography}{44}
\bibitem[{par()}]{parameters}
\bibinfo{note}{We use the parameters $\eta_z = 2\times10^{-3}$, $\lambda = 0.3
  \chi_{0}^{-1}$, $U \approx -2\times10^{7}\chi_{0}^{-1}$, and $W \approx
  2\times10^{12}\chi_{0}^{-1}$. The relative orders of magnitude between $C_s
  \sim 10^{2} \chi_{0}^{-1}$, $U$, and $W$ are associated with the equilibrium
  value of the orthorhombic distortion, $\delta \sim 10^{-3}$.}
\end{thebibliography}

\end{document}